# Estimation of atrial fibrillation from lead-I ECGs: Comparison with cardiologists and machine learning model (CurAlive), a clinical validation study

1st Necmettin Korucuk
*Department of Cardiology*
*Memorial Antalya Hospital*
Antalya, Turkey
necmettinmd@gmail.com

2nd Çağın Polat
*Chief Technical Officer*
*Notrino Research*
Ankara, Turkey
caginpolat@notrino.com

3rd Emine Selda Gündüz
*Vocational School of Health Services*
*Department of First and Emergency Aid*
*Akdeniz University*
Antalya, Turkey
seldagunduz@akdeniz.edu.tr

4th Onur Karaman
*Vocational School of Health Services*
*Department of Medical Imaging*
*Akdeniz University*
Antalya, Turkey
onurkaraman@akdeniz.edu.tr

5th Veysel Tosun
*Department of Cardiology*
*Şanlıurfa Education And Research Hospital*
Şanlıurfa, Turkey
veyseltosun8810@gmail.com

6th Mehdi Onaç
*Department of Cardiology*
*Akdeniz University Hospital*
Antalya, Turkey
mehdionac@gmail.com

7th Nezaket Yıldırım
*Faculty of Nursing, Department of Nursing Management,*
*Akdeniz University*
Antalya, Turkey
ozturknezaket@akdeniz.edu.tr

8th Yıldıray Çete
*Department of Emergency Medicine*
*Akdeniz University*
Antalya, Turkey
ycete@akdeniz.edu.tr

9th Kemal Polat
*Department of EEE*
*Bolu Abant Izzet Baysal University*
Bolu, Turkey
kpolat@ibu.edu.tr

*Abstract*—Electrocardiogram (ECG) recognition of cardiac arrhythmias is critical for cardiac abnormality diagnosis. Because of their strong prediction characteristics, artificial neural networks (ANNs) are the preferred method in medical diagnosis systems (MDS). This study presents a method to detect atrial fibrillation with lead-I ECGs using artificial intelligence. The aim of the study is to compare the accuracy of the diagnoses estimated by cardiologists and artificial intelligence over lead-I ECGs using 12-lead ECGs as references. To evaluate the performance of the proposed model, dataset were collected from China Physiological Signal Challenge 2018 (CPSC18).

In the study, diagnoses were examined in three groups as normal sinus rhythm (NSR), atrial fibrillation (AFIB) and OTHER. All rhythm and beat types except NSR and AFIB were labeled as OTHER super-class. OTHER contains First-degree atrioventricular blocks (AVBs), Conduction disturbances (CDs), Left bundle branch block (LBBB), Right bundle branch block (RBBB), Premature atrial contraction (PAC), Premature ventricular contraction (PVC), ST-segment depression (STD) and ST-segment elevated (STE) type ECGs. CurAlive A.I. model which is using DenseNet as a CNN architecture and continuous wavelet transform as feature extraction method, showed a great performance on classifying ECG's from only lead-I compared to cardiologists. The A.I. model reached the weighted average precision, recall, F1-score and total accuracy 94.1%, 93.6%, 93.7% and 93.6% respectively, and the average of each of the three cardiologists has reached weighted average precision, recall, F1-score and total accuracy 82.2%, 54.6%, 57.5% and 54.6% respectively. This study showed that the proposed CNN model CurAlive, can be used to accurately diagnose AFIB, NSR, and OTHER rhythm using lead-I ECGs to accelerate the early detection of AFIB as a cardiologist assistant. It is also able to identify patients into different risk groups as part of remote patient monitoring systems.

## I. Introduction

Atrial fibrillation (AFIB) is a form of cardiac arrhythmia characterized by the irregularity and speed of cardiac contractions (Kumar, Pachori, & Acharya, 2018). Atrial fibrillation (AFIB) is a common health problem not only in developed countries but also in developing countries (Hernandez et al., 2018; Xia et al., 2018; Wang, Wang, & Wang, 2020). AFIB affects some 90 million people worldwide (Wang, Wang, & Wang, 2020). According to estimates, 6–12 million people in the United States will be affected by this disease by 2050, and 17.9 million people in Europe will be affected by 2060. Other disorders including heart valve disease, diabetes mellitus, arterial hypertension and overweight/obesity, as well as metabolic syndrome and its components sleep apnea and inflammation, have significant epidemiological associations with AFIB. In addition to genetic factors, more recent research has centered on modifiable lifestyle factors such as alcohol intake and physical activity (Lippi, Sanchis-Gomar, & Cervellin, 2021).

Since AFIB can cause strokes, heart failure, and even death if not detected and treated promptly, automatic detection of AFIB is a critical requirement (Hernandez et al., 2018; Xia et al., 2018; Wang, Wang, & Wang, 2020). Furthermore, the economic and clinical burden of AFIB is significant and will continue to rise in the future due to high rates of hospitalization and extensive use of health resources (Asgari, Mehrnia, & Moussavi, 2015).

Some people may be unaware that they have atrial fibrillation because it does not often cause visible symptoms (Curry et al., 2018). Especially paroxysmal atrial fibrillation shows a silent course and its detection may require continuous monitoring. As a result, early detection and screening are important. Early detection of AFIB can thus allow for more effective management and avoid future complications (Mairesse et al., 2017).

ECG signals are often used to diagnose AFIB (Kumar, Pachori & Acharya, 2018; Wang, Wang, & Wang, 2020).

Electrocardiographic characteristics of AFIB on 12-lead standard ECG or single ECG not shorter than 30 seconds of AFIB including: Irregularly irregular R-R intervals (when atrioventricular conduction is not impaired), absence of distinct repeating P waves, and irregular atrial activations (Hindricks et al., 2021). In clinical settings, efficient and effective diagnosis of AFIB based on the ECG is useful but difficult (Xia et al., 2018). There is usually noise in the ECG signal obtained. As a result, highly accurate and robust detection of AFIB using lead-I, short-term ECG is useful but difficult (Nurmaini et al., 2020). There is a pressing need for an automated AFIB detection mechanism that can analyze large volumes of ECG data, accelerate diagnosis, and reduce health professionals workload (Wang, Wang, & Wang, 2020). Automatic AFIB diagnosis methods can be a alternative solution to meeting this requirement (Kumar, Pachori & Acharya, 2018; Wang, Wang, & Wang, 2020).

Artificial intelligence (A.I.) is a transformational technology that will affect all healthcare providers (McGrow, 2019). Machine learning is a branch of artificial intelligence that uses computational statistical algorithms to allow computers to learn from data without having to be specifically programmed (Mincholé & Rodriguez, 2019). Advances in neural networks, a machine learning methodology inspired by the biological learning process in the human brain, have been beneficial in healthcare (Clancy, 2020). Because of their strong prediction characteristics, artificial neural networks (ANNs) are the most commonly used method in medical diagnosis systems (MDS) (Özdemir & Danışman, 2011; Asgari, Mehrnia, & Moussavi, 2015; Smith et al., 2019). Especially, deep-learning algorithms can be used on large databases of electrocardiograms to detect irregular heart rhythms and mechanical dysfunction, which may help health professionals make better decisions (Mincholé & Rodriguez, 2019).

This study compares the performance of different cardiologists and an A.I. model called CurAlive with the reference of 12-lead gold standard ECG interpretations. The main aim of the study is to compare the performance of AFIB detection from only lead-I. Detecting only AFIB is not sufficient for automatic ECG interpretation. NSR and OTHER rhythm types should also be automatically interpreted. However, it is difficult to say that patient has NSR from only lead-I, because the ECGs can contain another confusing rhythm or beat types like AVBs, CDs, LBBB and RBBB.

## II. MATERIAL AND METHODS

### A. Clinical study procedure

The clinical study aimed to investigate how much information can be extracted from the ECG using only lead-I, and to compare the diagnostic performance of the artificial intelligence supported ECG analysis system of CurAlive, a product of Notrino Research, with the diagnostic performance of cardiologists on lead-I. Thus, the role of dry electrode and single lead mobile ECG devices in AFIB diagnosis will be examined in the future. The detailed schematic diagram of the clinical study is given in Figure 1.

In the study, lead-1 ECG recordings ranging from 10 seconds to 30 seconds were analyzed by 3 expert cardiologists and CurAlive. Cardiologists were asked to identify which of the ECG records were AFIB, NSR or OTHER. Cardiologists were instructed to choose the "OTHER" class if any abnormal rhythm or abnormal beat other than AFIB and NSR were involved. If there is another rhythm type at the same time as AFIB in the ECGs examined, this class was accepted as AFIB.

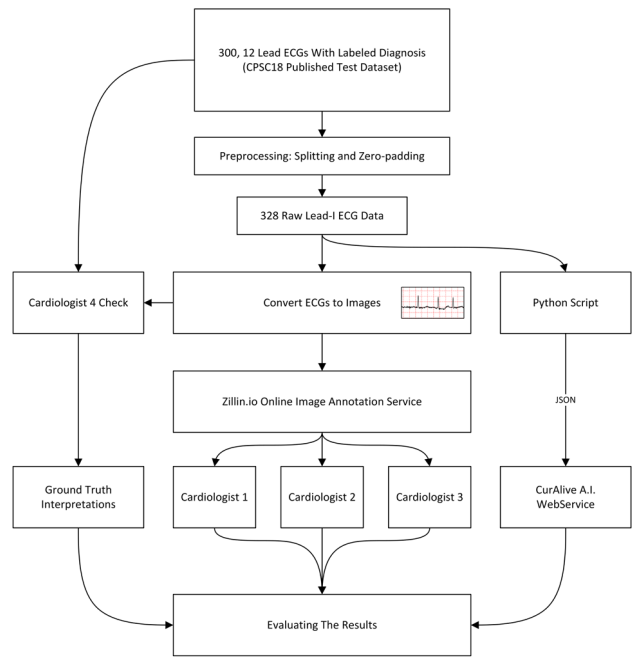

Fig. 1. The schematic diagram of the clinical study

The same ECG recordings were sent to CurAlive's artificial intelligence supported ECG analysis system running on the web and the results were recorded. CurAlive is an artificial intelligence supported mobile ECG patient tracking system that is in the development process. CurAlive consists of two different single lead and 6-lead ECG devices, artificial intelligence-supported ECG analysis system and a patient tracking system. At the same time, the devices can measure oxygen saturation level with embedded PPG sensors and can measure body temperature. The 6-lead ECG device also enables pacemakers to be monitored remotely with its pace-detection feature. The prepared ECG data were transmitted to the web-based and artificial intelligence supported ECG analysis system in JSON format with the username and password provided by the company in accordance with the data security rules and the results were obtained. CurAlive ECG analysis system is a system that can differentiate in NSR, AFIB, OTHER and NOISE classes by analyzing ECG signals between 10-30 seconds in time and frequency space using continuous wavelet transformation and convolutional neural networks. OTHER; It is a superclass used to identify ECG signals other than NSR and AFIB. NOISE; on the other hand, is a class that describes the noises that are not considered as ECG signals. CurAlive's CNN model was trained using PTB-XL, MIT-BIH Atrial Fibrillation Database and CurAlive ECG Noise Database.

Cardiologists and CurAlive have interpreted without knowing where the data came from and what's inside. Cardiologists also have evaluated ECGs independently from each other.

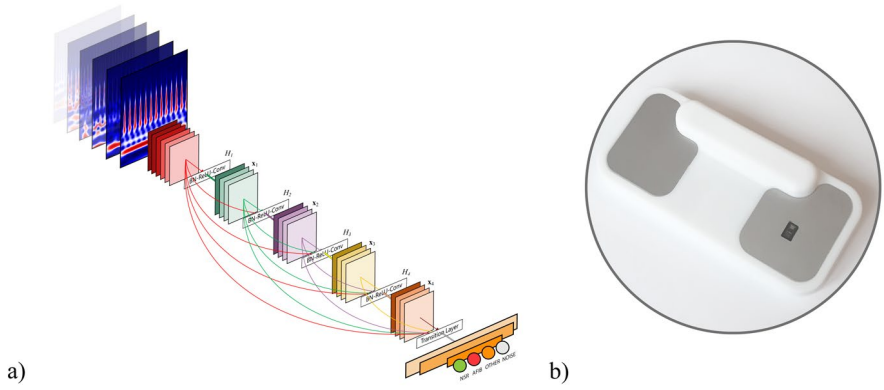

Fig. 2. a) CurAlive A.I. DenseNet type CNN architecture, b) CurAlive mobile ECG device

## B. Preparation of data and preprocessing ECGs

### 1) Database

300 12-lead ECG data with interpretations published by "The China Physiological Signal Challenge 2018 (CPSC18)" were used for testing purposes. The data set used in this study comparing the performance of cardiologists and CurAlive, includes 12-lead 300 ECG records. Inside the data set there are Atrial fibrillation (AFIB), Normal Sinus Rhythm (NSR), Atrioventricular blocks (AVBs), Conduction disturbances (CDs), Left bundle branch block (LBBB), Right bundle branch block (RBBB), Premature atrial contraction (PAC), Premature ventricular contraction (PVC), ST-segment depression (STD) and ST-segment elevated (STE) type heart rhythms and beats.

### 2) Pre-processing ECGs

The ECG data used in this research have been made suitable for use by being subjected to various pre-processes in order to be acceptable to web based CurAlive ECG analysis system.

All ECG signals except "NSR" and "AFIB" can only be identified in the "OTHER" class. The distribution of the grouped raw data set has been given in Figure 3.

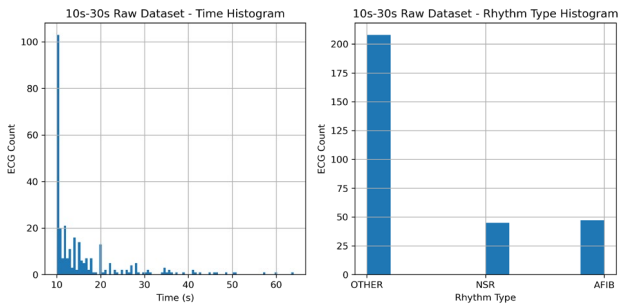

Fig. 3. Grouped raw data set. Left; ECG time distribution before splitting process, right; ECG class distributions before splitting process

Since 12-lead ECG records are gold standard, the diagnoses made by examining these records have been accepted as a reference. After dividing the ECGs into 10-30 second sub segments, 28 additional ECG signals have been generated. A fourth cardiologist was assigned to make sure that the ECG interpretations are the same with Chinese doctors and not changed after the splitting process. In this way, the diagnoses made by Chinese doctors were confirmed. In Figure 4 the splitted ECGs time and class distributions can be seen as histograms.

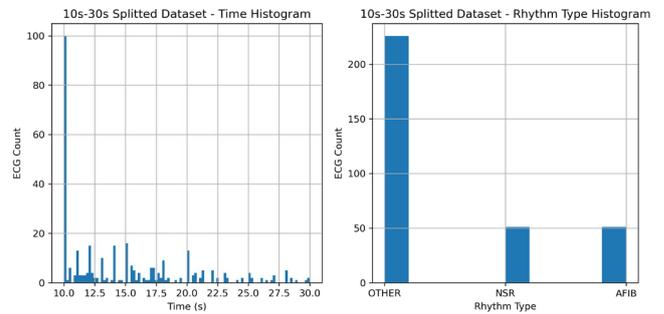

Fig. 4. Processed data set. Left; ECG time distribution after splitting process, right; ECG class distributions after splitting process

### 3) Lead-1 ECGs offered to cardiologists to diagnose

In order for cardiologists to examine ECGs, only the lead-I derivation of ECGs was selected from the 12-lead ECG data set and ECGs were recorded as pictures. The recorded pictures were uploaded to the web-based image annotation tool called "zillin.io", allowing each cardiologist to examine the same data separately and independently. The cardiologists, following an examination of the ECG results, selected the appropriate choice from the checkbox in the system, which included NSR, AFIB, OTHER, or NOT-SURE. Some ECG images presented to cardiologists have been given in Figure 5.

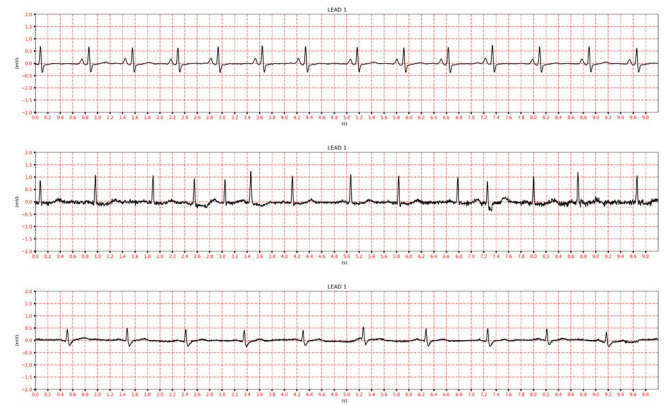

Fig. 5. Some of the ECGs uploaded to zillin.io online image annotation system.

## C. Statistical analysis

The results of the 12-lead ECGs were compared with the results of each cardiologist one by one and the confusion matrices were calculated. Precision, Recall and F1-Score values and weighted average values of each metric have calculated as given in the equations below.

$$Precision = \frac{true\ positives}{true\ positives + false\ positives} \quad (1)$$

$$Recall = \frac{true\ positives}{true\ positives + false\ negatives} \quad (2)$$

$$F_1 Score = \frac{2}{Precision^{-1} + Recall^{-1}} \quad (3)$$

$$Metric_{WA} = \frac{Metric_{AFIB} + Metric_{NSR} + Metric_{OTHER}}{N_{AFIB} + N_{NSR} + N_{OTHER}} \quad (4)$$

Kappa (κ) coefficients for interobserver agreement were assessed. κ coefficients are calculated with this formula below. In this formula Pr(a) represents the actual observed agreement, and Pr(e) represents chance agreement. The κ coefficient between CurAlive classifications and 12-lead reference ECG interpretation labels are calculated as 0.87. Kappa results are interpreted as follows: values ⩽ 0 as indicating no agreement and 0.01 – 0.20 as none to slight, 0.21 – 0.40 as fair, 0.41 – 0.60 as moderate, 0.61 – 0.80 as substantial, and 0.81–1.00 as almost perfect agreement (Landis and Koch, 1977; McHugh, 2012).

$$\kappa = \frac{P_r(a) - P_r(e)}{1 - P_r(e)} \quad (5)$$

95% CI intervals have also calculated. The intervals have an associated confidence levels that gives the probability with which the estimated interval will contain the true value of the parameter.

All metrics have been calculated for CurAlive and visualized on comparison tables. ROC curves (Receiver Operating Characteristics) and AUC (Area Under Curve) values were also calculated by using the estimated rhythms and estimation accuracy values.

## D. Internal consistency among cardiologists

The main reason for working with 3 cardiologists were to increase the reliability of the study. For this reason, it has desired to see how much collective decision that cardiologists make. Comparing the decisions of cardiologists with each other had been one way to demonstrate this. Therefore, 3 confusion matrices were prepared by comparing the decisions of each cardiologist with the decision of another cardiologist.

## III. RESULTS

### A. Evaluation of the model and performance comparison with the cardiologists

Using the image annotation tool zillin.io, all 328 ECG files prepared for cardiologists' evaluation were evaluated separately from one another according to the AFIB, NOT-SURE, NSR, and OTHER groups. All of the same ECGs were also analyzed by CurAlive for comparison.

The performance of each cardiologist and CurAlive have been tested by taking the evaluations made with 12-channel ECGs as a reference. Cardiologists have classified ECGs according to AFIB, NOT-SURE, NSR and OTHER classes. CurAlive is a model that can classify according to AFIB, NOISE, NSR and OTHER classes. In the data set consisting of 12-channel ECGs and labeled with reference diagnoses, only AFIB, NSR and OTHER classes have been available. As a result, macro average values of performance metrics have been ignored, as macro average values of performance metrics cannot produce accurate results.

In the initial evaluations, the weighted average precision, weighted average recall, weighted average F1-score, and accuracy metrics were compared for each cardiologist and CurAlive by using the result classification.

TABLE I.    DIAGNOSTIC PERFORMANCE COMPARISON: WEIGHTED PERFORMANCE METRICS

|  | *Weighted Avg. Precision* | *Weighted Avg. Recall* | *Weighted Avg. F1-Score* |
|---|---|---|---|
| **Cardiologist 1** | 82.9 | 67.1 | 70.8 |
| **Cardiologist 2** | 79.1 | 42.4 | 45.0 |
| **Cardiologist 3** | 84.6 | 54.3 | 56.6 |
| **Cardiologist Avg.** | 82.2 | 54.6 | 57.5 |
| **CurAlive** | **94.1** | **93.6** | **93.7** |

a. All metrics are given in percentage

In Table 1. the results of cardiologists and CurAlive's performance metrics have evaluated comparatively. The precision performance averages of the cardiologists participating in the study ranged from 79.1% to 84.6%. When the diagnostic performances performed by using the lead-I derivation were compared, the average precision of the diagnostics performed by the cardiologists was obtained 82.2%. CurAlive's average precision was found as 94.1%.

TABLE II.    DIAGNOSTIC PERFORMANCE COMPARISON: F1-SCORE PER CLASS AND TOTAL ACCURACY

|  | *F1-Score (AFIB)* | *F1-Score (NSR)* | *F1-Score (OTHER)* | *Accuracy* |
|---|---|---|---|---|
| **Cardiologist 1** | 81.2 | 54.4 | 72.1 | 67.1 |
| **Cardiologist 2** | 83.9 | 36.0 | 38.2 | 42.4 |
| **Cardiologist 3** | 90.2 | 42.0 | 52.3 | 54.3 |
| **Cardiologist Avg.** | 85.1 | 44.1 | 54.2 | 54.6 |
| **CurAlive** | **97.0** | **83.6** | **95.3** | **93.6** |

b. All metrics are given in percentage

The F1-Score per class and total accuracy diagnostic performance comparison of cardiologists and CurAlive was given in Table 2. The average F1 score of the cardiologists for AFIB was found to be 85.1%, whereas that of CurAlive was obtained as 97.0%. Moreover, while the average F1 score for NSR of cardiologists was determined as 44.1%, CurAlive's F1 score average was 83.6%. The average of OTHER F1-score was calculated as 54.2%, while CurAlive's average score was %95.3. It was determined that the accuracy of the cardiologists was 54.6%, while the accuracy of CurAlive was 93.6%.

TABLE III. EXAMINING CLASSIFICATION AGREEMENTS WITH 12-LEAD REFERENCE ECG INTERPRETATIONS (COHEN'S KAPPA STATISTIC)

| | κ Value | Standard Error | 95% CI | p Value |
|---|---|---|---|---|
| Cardiologist 1 | 0.47 | 0.039 | 0.39 to 0.55 | .000* |
| Cardiologist 2 | 0.24 | 0.029 | 0.18 to 0.30 | .000* |
| Cardiologist 3 | 0.35 | 0.033 | 0.29 to 0.41 | .000* |
| CurAlive | **0.87** | **0.028** | **0.81 to 0.92** | .000* |

c. * p<.001

Table 3 shows the results of the examination of the 12-channel reference ECG interpretations and classification agreements. According to the results of the K calculations, there is a perfect match between the 12-lead gold standard ECG interpretations and CurAlive. Two cardiologists showed fair agreements and one cardiologist showed a moderate agreement between 12-lead reference interpretations.

True positive, false positive, true negative and false negative values are extracted from calculated confusion matrices. These matrices also gave us the ability to evaluate the results in a subjective way. On these matrices, it can be seen that CurAlive made predictions that are almost perfect compared to cardiologists. Cardiologists confused OTHER class as NSR most of the time. This is expected because without any useful information from other ECG derivations, only lead-I derivation is a poor way to interpret ECGs with humans.

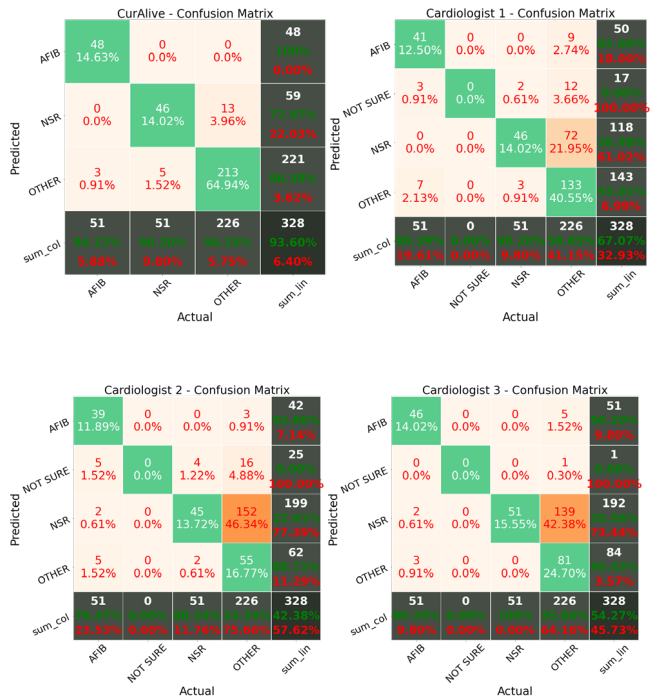

Fig. 6. Confusion matrices between ground truth 12-lead interpretations and each cardiologist and CurAlive

In Figure 6. Three cardiologists and CurAlive's Confusion Matrix have given in the comparison made with reference to 12-lead ground truth interpretations. The number of ECGs on which cardiologists are unsure is extremely low. The irregularity of R waves and the absence of P waves are two important diagnostic parameters when diagnosing AFIB. The fact that P waves in lead-I appear flatter than lead-II, which makes AFIB diagnosis from lead-I difficult. However, since R waves can be seen easily in lead-I, more accurate results can be seen in AFIB diagnosis compared to other classes. It appears that all three cardiologists have made serious mistakes while interpreting the NSR and OTHER classes. The common occurrence of errors demonstrates the strength of diagnostic correlation between cardiologists. The main reason for this is that ECG records belonging to rhythms and beats of AVB, CDs, LBBB, RBBB, PAC, PVC, STD and STE types in OTHER class are difficult to classify by using only lead-I derivation. NSR and other specified rhythm and beat types can create very similar patterns in lead-I derivation.

When the CurAlive's confusion matrix, shown in Figure 6, is analyzed, it can be shown that the model has high selectivity across all groups. It is able to detect rhythms and beats of AVB, CDs, LBBB, RBBB, PAC, PVC, STD and STE types even from lead-I and distinguish them from NSR class. In AFIB detection, it is seen that it can classify AFIB from lead-I with a high performance with 100% precision, 96.1% recall and 98.0% F1-Score value.

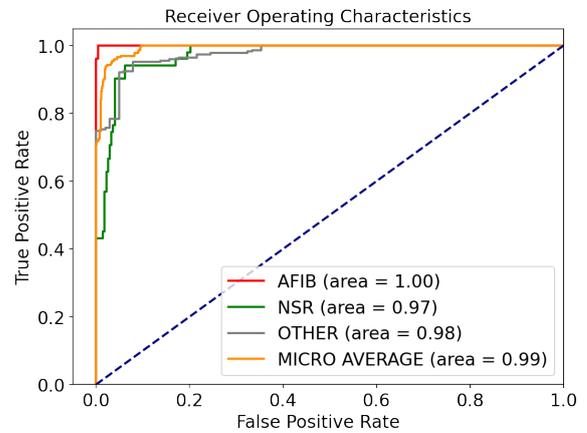

Fig. 7. CurAlive receiver operating characteristics curves and area under curve values

The trained A.I. system was evaluated on the CPSC18 test dataset. We used the receiver operating characteristic (ROC) curves to evaluate the diagnostic accuracy. On the test dataset, the ROC curve (Fig. 7) showed AUC of three categories were respectively 1.00 (for AFIB), 0.95 (for NSR), 0.97 (for OTHER), and the micro average is 0.99. Besides, precision and recall for AFIB were 96.1% and 98.0% respectively. The AUC value for AFIB calculated as 1.00 that means the model has perfect diagnostic accuracy for diagnosing AFIB.

*B. Total agreement between cardiologists*

All cardiologists have given similar performance for the AFIB class. But it can easily be seen in the confusion matrices between cardiologists that there are some misclassifications for AFIB class (Figure 8). It is due to the absence of P waves that can't be easily seen on lead-I ECGs. On the other hand NSR and OTHER classes are mostly confused with each other. From the results the p values are calculated as zero that means there is no significant correlation between cardiologists' decisions. With this knowledge it can be said that when diagnosing most of the OTHER type ECGs the cardiologists made near random choices with no correlation. So for interpreting some ECGs the lead-I is most of the time useless and difficult for humans for making decisions about the ECGs.

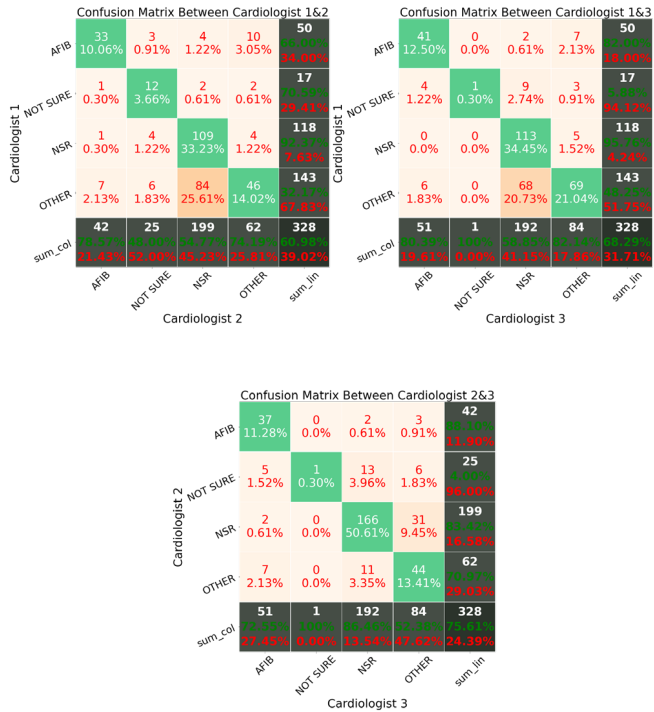

Fig. 8. Confusion matrices between each cardiologists

## IV. DISCUSSION

Atrial fibrillation is mostly asymptomatic, making it difficult to diagnose, but it is associated to stroke, heart failure, and death. Existing screening methods necessitate long-term monitoring and are cost and yield constrained. This study is to compare the accuracy of the diagnoses estimated by doctors and artificial intelligence over lead-I ECGs using 12-lead ECGs as ground truth references. In the literature, many studies on AFIB diagnosis using various artificial intelligence models have been found.

The development of an automated and efficient method that monitors AFIB in real time will likely be a key component in achieving public health goals for reducing disease-induced deaths (Faust, Ciaccio., & Acharya, 2020). In a study conducted by Duerte et al. (2019), it has been suggested that lead-I ECG can be used as an easily accessible and economical pioneering evidence in the diagnosis of AFIB in "primary care" healthcare institutions (Duarte et al., 2019). In another study reported by Taggar et. al. (2015), three studies were undertaken: 1) a systematic review and meta-analysis of the diagnostic accuracy of methods for detecting pulse irregularities caused by AFIB, 2) a systematic review and meta-analysis of the diagnostic accuracy of methods for diagnosing AFIB using 12-lead ECG and 3) a survey of HCPs in primary care about screening implementation. Automated ECG-interpretation software most accurately excluded AFIB, although its ability to diagnose this was similar to all healthcare professionals. Indirect comparisons with software found healthcare professionals (Taggar et.al., 2015). In Pourbabaee et al.'s (2018) study, 10-second ECG data of 33 women were taken and evaluated with CNN. ECGs were monitored by the patients at least 5 times a day for six weeks. On validation and testing data sets created from different recordings of the same participants, an overall window accuracy of 95.25% and 95.95% achieved (Pourbabaee et al., 2018).

Asensi et al. (2020), shows that automated AFIB detection using a heart rhythm monitor with sensors and algorithms that analyze the PPG signal and the electrocardiographic signal corresponding to lead-I is possible and accurate. The AFIB detection algorithm based on the PPG signal had a sensitivity of 91% and a specificity of 96% (diagnostic accuracy: 93%). The automatic algorithm based on the electrocardiographic signal had a sensitivity of 94% and a specificity of 96% (diagnostic accuracy: 95%) (Asensi et al., 2020). Boriani et al. (2021), a total of 2814 subjects (median age 68 years) underwent AFIB screening by a handheld single-lead ECG device. Overall, 56 subjects (2.0%) were diagnosed with AFIB, as a result of 12-lead ECG following a positive/suspected recording. Screening identified AFIB in 2.9% of the subjects ≥ 65 years. In addition, they found sensitivity (94.4%) using 12 lead ECG data (Boriani et. al.,2021).

In a study conducted by Welton et. al. (2019), it has been suggested that 12-lead ECG had a sensitivity of 93% and a specificity of 97%; individual studies in the review had a sensitivity of 68% to 100% and a specificity of 76% to 100% (Welton et al., 2017). Liu et al. (2018), it was controlled using 12-lead ECG data from 60 normal human and 50 AFIB patients and achieved 88% sensitivity and 75% specificity (Liu et al., 2018). Limam and Precioso (2017) proposed a system to classify ECG signals into AFIB, Normal, Alternative rhythm, or noisy signal using six layers of CRNN, including two separate CNNs to extract similar patterns, one from heart rate and the other from ECG, which are combined into an RNN classifier to reflect the progression of the removed patterns. A Support Vector Machine (SVM) with a 77% accuracy is used to access the final result (Limam & Precioso, 2017).

By searching at other studies in the literature, it's clear that single lead, 6-Lead, and 12-Lead ECGs, as well as evaluations made with various data sources, yield results with varying confidence intervals. In this research, as a result of the evaluation of the artificial intelligence model called CurAlive, it was found as 94.1% precision, 93.6% recall and 93.7% F1-Score. CurAlive has found that its diagnostic power is significantly higher than cardiologists when compared with cardiologists and the actual ground truth diagnosis of 12 Lead ECGs. In addition, obtaining results with an accuracy close to 12-lead ECGs using only lead-I ECGs is a striking result of this research.

Atrial fibrillation burden may increase by >60% in 2050 (Lippi et.al.,2019). AFIB is one of the most important global health problem. Early AFIB diagnosis has a very important role of solving this global problem. Thus, mobile ECG patient tracking systems and A.I. powered decision making systems are going to play a critical role. This clinical study showed that CurAlive is a reliable, accessible and inexpensive tool to make possible the early diagnosis of AFIB.

## V. CONCLUSION

Lead-I ECG devices have frequently been used in the diagnosis of normal sinus rhythm and atrial fibrillation. It is not always easy for healthcare professionals to diagnose normal sinus rhythm from lead-I ECGs. Because some rhythms can be interpreted differently in lead-I derivation, it is recommended in the literature to be controlled with 12-lead ECGs. However, it is important to develop the systems using lead-I ECGs due to reasons such as limited accessibility of 12-

lead ECG devices. Also 12 lead devices are impractical, expensive and incompatible with remote monitoring systems with existing technology. CureAlive identified AFIB using lead-I ECGs, with nearly 12-lead ECGs sensitivity and significantly higher diagnostic power than cardiologists which use lead-I.

These results are promising in increasing the early diagnosis capacity of remote patient monitoring systems. In this way, it can be said that CureAlive can contribute to the diagnosis of cardiac emergencies practically, early, quickly and cheaply. With future enhancements, it is estimated that CurAlive will also be able to detect other abnormal rhythms and beats that can only be identified with 12-lead ECGs using lead-I ECGs. Based on these results, it is planned to conduct a research on diagnosis of MI from lead-I ECGs using prospective patient data. It can be speculated that CurAlive will be capable of diagnosing other abnormalities thanks to its flexible design, so it is likely to be used for the diagnosis of other heart, kidney, and endocrinological diseases in the early-future.

## VI. ACKNOWLEDGMENT

We would like to thank Notrino Research for their help in this study and for allowing us to use the artificial intelligence model they developed. We would also like to thank Dr. Okay Işlak for statistical contributions.